\documentclass[11pt]{article}
\usepackage[margin=1in]{geometry}
\usepackage{amsmath,amssymb,amsthm}
\usepackage{hyperref}

\newtheorem{theorem}{Theorem}
\newtheorem{remark}{Remark}
\newtheorem{definition}{Definition}

\title{A Note on Non-Composability of Layerwise Approximate Verification for Neural Inference}
\author{Or Zamir \\ Tel Aviv University}
\date{}

\begin{document}
\maketitle

\begin{abstract}
A natural and informal approach to verifiable (or zero-knowledge) ML inference over floating-point data is:
``prove that each layer was computed correctly up to tolerance $\delta$; therefore the final output is a reasonable inference result''.
This short note gives a simple counterexample showing that this inference is false in general:
for any neural network, we can construct a functionally equivalent network for which adversarially chosen
approximation-magnitude errors in individual layer computations suffice to steer the final output arbitrarily
(within a prescribed bounded range).
\end{abstract}

\section{Introduction}

Recent work of Bitan, DeStefano, Goldwasser, Ishai, Kalai, and Thaler~\cite{BDGIKT25} develops an
\emph{approximate} variant of the classical sum-check protocol that natively reasons about approximate arithmetic.
In their framework, the prover and verifier exchange approximate real/complex numbers, each round replaces exact equalities by comparisons up to a tunable tolerance~$\delta$, and soundness degrades reasonably with respect to~$\delta$. Their results yield completeness, soundness, and round-by-round soundness for this approximate protocol, and are based on analytic control of low-degree
polynomials rather than arithmetization or code-based arguments. A conceptual takeaway is a ``black-box feasibility'' result for verifying approximate arithmetic: the protocol does not depend on the bit-level representation used to implement approximate operations, only on satisfying stated error bounds.

A central motivation in~\cite{BDGIKT25} is the mismatch between exact verifiable computation and numerical workloads,
including machine learning: floating-point computations are inherently approximate, can be non-associative, and may be
non-deterministic on modern hardware, yet exact proof systems typically certify exact execution of a computation over finite fields or integers. As emphasized in~\cite{BDGIKT25}, the prevailing approach in proof systems for numerical tasks
is to encode approximate real-valued computations as exact arithmetic over finite fields, which can incur substantial overhead and distort the intended computation. This issue has been
especially visible in the growing literature on \emph{zk-ML} (verifiable/zero-knowledge ML inference), including systems
and applications such as~\cite{KDZ24,Lab24,GGPZ17} (see also the discussion and references in~\cite{BDGIKT25}).

It is therefore tempting to hope for a very simple route to ``native-precision'' zk-ML: given an ML model expressed as a
sequence of floating-point (or real-valued) layers, one proves \emph{layer by layer} that the claimed intermediate vector
is within tolerance~$\delta$ of the honest layer output, and then concludes that the final output is a reasonable inference
result. The purpose of this note is to rule out such a \emph{generic black-box composition} statement. Namely, even for
ReLU networks with bounded weights, layerwise $\delta$-consistency permits adversarial transcripts whose errors amplify
across layers and can be used to steer the final output essentially arbitrarily within a prescribed range. This does not
contradict~\cite{BDGIKT25}, whose soundness analysis is tailored to sum-check and explicitly controls how relaxed checks
interact across rounds; rather, it highlights that without additional stability assumptions or protocol-level reasoning
about error propagation, local approximate checks do not compose.

Our observation is that this simple approach to zk-ML fails in a very strong sense: every neural network has a functionally equivalent network in which injecting tiny, adversarially chosen errors into individual layer computations is enough to drive the final output to an arbitrary value.
While it is well known that the numerical stability of multi-layer networks can degrade exponentially with depth (e.g.,~\cite{yuan2025understanding,budzinskiy2025numerical}), one might hope that ``reasonable'' networks (those representing natural functions) avoid such worst-case error accumulation. Our construction rules out this optimism.

Moreover, we highlight that in realistic zk-ML threat models one should not assume the network is ``natural''. Typically, the network is constructed by an adversary and then committed to, perhaps alongside a proof that it passed certain tests or audits (which, in practice today, would largely amount to black-box benchmark evaluations). 
Thus, if the adversary can preserve the network’s functionality exactly (so it still passes auditing) while making it susceptible to later attacks on the zk-ML protocol, it has every incentive to do so.

\section{Model: layerwise approximate verification}

Fix a depth-$k$ feed-forward network with pointwise activation $\sigma$ (e.g.\ ReLU).
For weight matrices $A_1,\dots,A_{k-1}$ and a final linear map $A_k$, define exact inference by
\[
y_0 := x,\qquad y_i := \sigma(A_i y_{i-1})\ (i=1,\dots,k-1),\qquad y_k := A_k y_{k-1}.
\]
Write $F(x):=y_k$.

A ``line-by-line approximate verifier'' typically receives from the prover a transcript
$\widehat{y}_0,\widehat{y}_1,\dots,\widehat{y}_k$ and checks that each transition is correct up to error
$\delta$ with respect to the \emph{prover-supplied previous state} (so any per-step deviation is fed into
subsequent checks).

\begin{definition}[Layerwise $\delta$-consistency]
Fix a norm $\|\cdot\|$ (below we use $\ell_\infty$).
A transcript $(\widehat{y}_0,\dots,\widehat{y}_k)$ is \emph{$\delta$-consistent} for $(A_1,\dots,A_k,\sigma)$
and input $x$ if $\widehat{y}_0=x$ and for all $i=1,\dots,k-1$,
\[
\|\widehat{y}_i - \sigma(A_i \widehat{y}_{i-1})\|_\infty \le \delta,
\]
and also
\[
\|\widehat{y}_k - A_k \widehat{y}_{k-1}\|_\infty \le \delta.
\]
\end{definition}

\noindent
When $\delta=0$ this pins down a unique transcript (so local checks compose, a phenomenon commonly used for integral or finite field computations). For $\delta>0$, acceptance
means only that the output is reachable by a perturbed dynamical system where the prover can choose the
perturbations adversarially.

\section{Construction: exact equivalence, arbitrary approximate outputs}

We show that for any given network $F$ one can build a \emph{functionally identical} network $F'$ such that
layerwise $\delta$-consistency no longer implies anything meaningful about the final output: the prover can
steer the output to essentially any target in a bounded range.

\begin{theorem}[Universal steering under layerwise $\delta$-checks]\label{thm:steer}
Fix the activation $\sigma=\mathrm{ReLU}$ on hidden layers and let the final layer be linear. Fix any network
$F=(A_1,\dots,A_k)$ with width~$\bar{m}$, output dimension $m$, weight bound~$g>1$, output bound $R>0$,
and any tolerance $\delta>0$. Here \emph{output bound $R$} means that for all inputs $x$ under consideration,
$\|F(x)\|_\infty \le R$. Assume \emph{weight bound $g$} means that every entry of every $A_i$ has absolute value at most $g$.
Then there exists a depth-$k$ network $F'=(A'_1,\dots,A'_k)$ with width~$\bar{m}+O(m)$, output dimension~$m$,
and weight bound~$\max\!\left(g,\frac{2R}{\delta g^{k-2}}\right)$ such that:

\begin{enumerate}
\item \textbf{(Exact equivalence)} For all inputs $x$, exact inference satisfies $F'(x)=F(x)$.
\item \textbf{(Steering)} For every input $x$ and every target $z\in\mathbb{R}^m$ with
$\|z\|_\infty\le R$, there exists a $\delta$-consistent transcript for $F'$ on input $x$ whose final output
equals $z$.
\end{enumerate}
Consequently, verifying each layer ``up to $\delta$'' does \emph{not} imply that the verified output is close
to the true inference $F(x)$, even when the verified model is exactly functionally equivalent to the intended model.
\end{theorem}

\paragraph{Construction.}
Let $F=(A_1,\dots,A_k)$ be a depth-$k$ ReLU network of width $\bar m$ and output dimension $m$.
Assume $F$ has weight bound $g>1$, i.e.\ every entry of every $A_i$ has absolute value at most $g$.
We define $F'=(A'_1,\dots,A'_k)$ by widening each hidden layer by an auxiliary \emph{trigger channel}
$(t^+,t^-)\in\mathbb{R}_{\ge 0}^m\times\mathbb{R}_{\ge 0}^m$, adding $2m$ coordinates.
Thus the width becomes $\bar m + 2m = \bar m + O(m)$.

\smallskip
\noindent\textbf{Layer 1.}  Define
\[
A'_1 :=
\begin{pmatrix}
A_1\\
0\\
0
\end{pmatrix},
\qquad\text{so}\qquad
\mathrm{ReLU}(A'_1 x)=\big(\mathrm{ReLU}(A_1x),\,0,\,0\big).
\]

\smallskip
\noindent\textbf{Hidden layers $i=2,\dots,k-1$.}
Define the block-diagonal map
\[
A'_i :=
\begin{pmatrix}
A_i & 0 & 0\\
0 & g I_m & 0\\
0 & 0 & g I_m
\end{pmatrix}.
\]
Hence, if $t^\pm\ge 0$, then ReLU leaves them nonnegative and they get multiplied by $g$ each layer:
$(t^+,t^-)\mapsto (g t^+, g t^-)$.

\smallskip
\noindent\textbf{Final layer.}
Let the final layer be linear, and define
\[
A'_k := \big( A_k \ \ \ M I_m \ \ -M I_m \big),
\qquad\text{where}\qquad
M := \frac{2R}{\delta g^{k-2}}.
\]
Thus the output of $F'$ on a hidden state $(y,t^+,t^-)$ equals
\[
A_k y + M(t^+-t^-).
\]

\smallskip
\noindent\textbf{Weight bound.}
By construction, all new weights are in $\{0,1,g,M\}$ up to sign, so the weight bound of $F'$
is $\max(g,M)=\max\!\left(g,\frac{2R}{\delta g^{k-2}}\right)$, as claimed.

\paragraph{Proof of Theorem~\ref{thm:steer}.}
\emph{Exact equivalence.}
Under exact inference, the auxiliary coordinates after layer 1 are $(t^+,t^-)=(0,0)$.
The hidden-layer update multiplies these by $g$ but keeps them at $0$ forever, and the final layer adds
$M(t^+-t^-)=0$. Therefore $F'(x)=F(x)$ for all $x$.

\smallskip
\emph{Steering.}
Fix an input $x$ and let $y^*:=F(x)\in\mathbb{R}^m$ denote the true output of $F$ (and thus of $F'$ under
exact inference). Fix any target $z\in\mathbb{R}^m$ with $\|z\|_\infty\le R$.
Define $\Delta := z-y^*$. By the assumed output bound $\|y^*\|_\infty\le R$, we have
\[
\|\Delta\|_\infty \le \|z\|_\infty + \|y^*\|_\infty \le 2R.
\]
Let $T:=g^{k-2}$ denote the trigger-channel amplification from layers $2$ through $k-1$.
We will inject a small trigger at layer 1 so that after amplification and the final linear combination it
produces the desired shift $\Delta$.

For a vector $v\in\mathbb{R}^m$, let $(v)_+$ denote its coordinatewise positive part: $((v)_+)_j = \max\{v_j,0\}$.
Define nonnegative vectors
\[
u^+ := \frac{(\Delta)_+}{M T},\qquad u^- := \frac{(-\Delta)_+}{M T}.
\]
Then $u^\pm\ge 0$ and
\[
M T (u^+ - u^-) = \Delta.
\]
Moreover, since $M=\frac{2R}{\delta g^{k-2}}$ and $T=g^{k-2}$, we have $MT = \frac{2R}{\delta}$, hence
\[
\|u^\pm\|_\infty \le \frac{\|\Delta\|_\infty}{MT} \le \frac{2R}{2R/(\delta)} = \delta.
\]

Now we provide the prover transcript. Let $\widehat{y}_0:=x$. For layer 1, the honest output would be
$(\mathrm{ReLU}(A_1x),0,0)$; instead the prover sends
\[
\widehat{y}_1 := \big(\mathrm{ReLU}(A_1x),\,u^+,\,u^-\big).
\]
This is $\delta$-consistent for layer 1 since the deviation from the honest layer output occurs only in the
auxiliary coordinates, with $\ell_\infty$ magnitude at most $\delta$.

For each hidden layer $i=2,\dots,k-1$, let $\widehat{y}_i$ be the \emph{exact} evaluation of layer $i$ of $F'$
on input $\widehat{y}_{i-1}$. Since the transcript follows the true transition on the already-perturbed state,
each such step has zero additional error, and in particular remains $\delta$-consistent.

At the end of the hidden layers, the trigger channel equals $(T u^+, T u^-)$, and thus the final output is
\[
\widehat{y}_k
= A_k(\cdot) + M(Tu^+ - Tu^-)
= y^* + \Delta
= z.
\]
Therefore there exists a $\delta$-consistent transcript whose final output is the arbitrary target $z$.
\hfill$\square$

\begin{remark}[Parameter choices]
The required ``steering weight'' is $M=2R/(\delta g^{k-2})$, i.e., amplification by $g$ across depth reduces the
needed final gain.
For a concrete sanity check: taking $R\approx 20$ (logit scale), $\delta\approx 10^{-3}$ (FP16-scale absolute tolerance),
$g=2$, and $k=20$ gives $M \approx 40/(10^{-3}\cdot 2^{18}) \approx 0.15$, which are completely standard-sized weights.
Thus, even tolerances comparable to floating-point rounding errors can be sufficient for adversarial steering under
layerwise approximate verification.
\end{remark}

\section{Takeaway}
Our most direct takeaway is that layerwise ``approximate correctness'' is not composable, even in standard-architecture neural networks computing natural functions. 
Without additional assumptions or protocol-level reasoning about error propagation, it does not
certify that the final output is close to the true inference result.

Further, we highlight that zk-ML protocols should not be designed while assuming that the network is natural or was constructed innocently, only that it \emph{appears} so to whatever audits or certifications are applied.

\end{document}